\documentstyle[aps,12pt,manuscript]{revtex}
\begin{document}
\preprint{INJE--TP--95--2}
\def\overlay#1#2{\setbox0=\hbox{#1}\setbox1=\hbox to \wd0{\hss #2\hss}#1%
\hskip -2\wd0\copy1}

\title{ Blueshift of tachyon in the charged 2D black hole}

\author{ H.W. Lee and Y. S. Myung }
\address{Department of Physics, Inje University, Kimhae 621-749, Korea}
\author{Jin Young Kim}
\address{Division of Basic Science, Dongseo University, Pusan 616-010, Korea}

\maketitle

\vskip 1.5in

\begin{abstract}
We study the propagation  of string fields
 (metric $G_{\mu\nu}$, Mawxell gauge potential $A_{\mu}$, dilaton $\Phi$,
and tachyon $T$) in
a two-dimensional (2D) charged black hole.
It is shown that the tachyon is a propagating field  both inside  and
outside the black hole.
This becomes infinitely blueshifted at the inner horizon.
We confirm that the inner horizon is unstable, whereas the outer horizon is
stable.
\end{abstract}

\vskip .5in
PACS number(s) : 04.70.Bw, 11.25 Db, 11.55.Ds
\newpage

Lower dimensional theories of gravity provide  a simplified context in
which to study
black hole physics [1].
The non-triviality of these models
arises from the non-minimal coupling of the dilaton to the scalar curvature.
A dilaton potential of the type produced by the string loop corrections may
induce multiple
horizons [2].   For example, the 2D charged black hole from heterotic
string theories has shown this
feature. This has many analogies with  the Reissner-Nordstr\"om black hole in
 4D general relativity.
In addition to the event (outer) horizon ($r_+$), there exist
the Cauchy (inner) horizon ($r_- $) in both the 2D charged and
Reissner-Nordstr\"om black hole.

Penrose [3]  has first pointed out that the Cauchy horizon of the
Kerr-Newman black hole
in 4D gravity is unstable due to the infinite blueshift  of the infalling
radiation.
Here the large blueshift of infalling matter means a general divergence of
the field energy density
as the evolution approaches the Cauchy horizon.
McNarama [4] has demonstrated that a test scalar field evolved to have
unbounded enegy density on
$r_-$. The literatures in [5-8] considered the stability problem of the
Cauchy horizon within
the Reissner-Nordstr\"om geometry. These indicated that the Cauchy horizon
is unstable to the external perturbations.
On the other hand, Poisson and Israel [9] showed that the mass parameter
near $r_-$ becomes
unbounded only when both the infalling and outgoing radiations are present.
This is called the  mass inflation.
Ori [10] confirmed this mass inflation within a simplified model.
Recently, Husain [11] showed that the mass inflation also occurs in the 3D
space-time.
More recently, Chan and Mann [12] investigated the same problem in the 2D
charged dilaton gravity.
In this case the stress-energy tensor for matter is taken as a null fluid.

 This paper is concentrated on the study of the propagation  of string fields
 (metric $G_{\mu\nu}$, Mawxell gauge potential $A_{\mu}$, dilaton $\Phi$,
and tachyon $T$) in
the 2D charged black hole. We will show that the  tachyon is the only
propagating field in the background
of double horizons.
Following Ref.[6], we use the tachyon to investigate
inner and outer horizons instead of a null fluid matter.
Furthermore, the region outside  the black hole is also  studied when the
charge of
the black hole vanishes ($Q=0$).

{}From the conformal invariance of the heterotic 2D string theories, one can
derive
the $\beta$-function equations [2,13,14]

\begin{equation}
R_{\mu\nu} + 2 \nabla_\mu \nabla_\nu \Phi
-F_{\mu \rho} F_\nu^{~\rho} - {1\over 2} \nabla_\mu T \nabla_\nu T = 0,
\end{equation}

\begin{equation}
\nabla^2 \Phi - 2 (\nabla \Phi)^2 + {1 \over 2} \alpha^2 + {1 \over 4} F^2
+ {1 \over 2} T^2  = 0,
\end{equation}

\begin{equation}
 \nabla_\mu F^{\mu \nu} - 2 (\nabla_\mu \Phi) F^{\mu \nu} = 0,
\end{equation}

\begin{equation}
\nabla^2 T - 2 \nabla_\mu \Phi \nabla^\mu T + 2 T = 0,
\end{equation}
where $ F_{\mu \nu } = \partial_{[\mu} A_{\nu]}$ is the Maxwell field.
The above equations are also derived from the requirement that the fields
must be an extremum
of the low-energy string action [14]

\begin{equation}
S_{l-e} = \int d^2 x \sqrt{-G} e^{-2\Phi}
   \big \{ R + 4 (\nabla \Phi)^2 + \alpha^2 - {1 \over 2} F^2 - {1 \over 2}
(\nabla T)^2 +
 T^2 \big \}.
\end{equation}
For an example, we consider the variation of $S_{l-e}$ with respect to the
metric $G_{\mu\nu}$.
This leads to $T_{\mu\nu} = 0$ with  the stress-energy tensor
$$T_{\mu\nu}= T^{\Phi}_{\mu\nu} + T^t_{\mu\nu} + T^F_{\mu\nu}
= - 2 \nabla_\mu \nabla_\nu \Phi + 2 G_{\mu\nu}(
\nabla^2 \Phi - (\nabla \Phi)^2 + 2) $$
\begin{equation}
 + { 1\over 2}\nabla_\mu T\nabla_\nu T + {1 \over 4}G_{\mu\nu}
(2T^2 - (\nabla T)^2 ) + F_{\mu\rho}F_{\nu}~^{\rho} - {1 \over 4}G_{\mu\nu}F^2.
\end{equation}
For convenience, let us take the following transformation
\begin{equation}
-2\Phi \rightarrow \Phi,~~~ T \rightarrow \sqrt 2 T, ~~~-R  \rightarrow R.
\end{equation}
Then the equations of motion are given by
\begin{eqnarray}
&&R_{\mu\nu} + \nabla_\mu \nabla_\nu \Phi + F_{\mu\rho}F_{\nu}^{~\rho} +
\nabla_\mu T \nabla_\nu T = 0,  \\
&& (\nabla \Phi)^2 +  \nabla^2 \Phi - {1 \over 2} F^2  - 2 T^2 - 8 = 0,  \\
&&\nabla_\mu F^{\mu \nu} + (\nabla_\mu \Phi) F^{\mu \nu} = 0,   \\
&&\nabla^2 T + \nabla \Phi \nabla T + 2 T = 0,
\end{eqnarray}
where we set $\alpha^2 = 8$.
The equation (9) comes from $T_{\mu}^{~\mu} =0$ with the substitution (7).
For a later purpose,
we have the stress-energy tensor for tachyon (after substitution of (7))
\begin{equation}
 T^t_{\mu\nu}= \nabla_\mu T\nabla_\nu T + {1 \over 2}G_{\mu\nu}
(2T^2 - (\nabla T)^2 ).
\end{equation}
The charged black hole solution to the above equations is given by setting
the tachyon ($T$) to
zero [2]

\begin{equation}
\bar \Phi = 2 \sqrt 2 r,~~~ \bar F_{tr} = Q e^{-2 \sqrt 2 r},~~~ \bar T = 0,
{}~~~ \bar G_{\mu\nu} =
 \left(  \begin{array}{cc} - f & 0  \\
                             0 & f^{-1}   \end{array}   \right) ,
\end{equation}
with
\begin{equation}
f = 1 -  e^{- 2 \sqrt 2 r} + {Q^2 \over 8}e^{- 4 \sqrt 2 r}.
\end{equation}
Introducing the new coordinate $y =\exp(2\sqrt2 r)$, $f={1 \over y^2}(y^2 -
y +{Q^2 \over 8})$.
{}From $f=0$, the double horizons ($r_{\pm}$) are given by
\begin{equation}
r_{\pm} = { 1 \over 2 \sqrt 2 }\log y_{\pm}
\end{equation}
with
\begin{equation}
y_{\pm}= { {1 \pm  \sqrt { 1 - {Q^2 \over 2}}} \over 2}.
\end{equation}
Here $r_{+}(r_{-})$ correspond to the outer (inner) horizons. In the case
of $Q =1$, we have
$r_+ = - 0.056$ and $r_- = - 0.679$.

To study the propagation of string fields, we introduce small perturbation
fields  around
the background solution as [15]
\begin{eqnarray}
&&F_{tr} = \bar F_{tr} + {\cal F}_{tr} = \bar F_{tr} [1 - {{\cal F}(r,t)
\over Q}],        \\
&&\Phi = \bar \Phi + \phi(r,t),                       \\
&&G_{\mu\nu} = \bar G_{\mu\nu} + h_{\mu\nu}  = \bar G_{\mu\nu} [1 - h
(r,t)],     \\
&&T = \bar T + \tilde t \equiv \exp (-{\bar \Phi \over 2}) [ 0 + t (r,t) ].
\end{eqnarray}
where we choose the metric perturbation ($h_{\mu\nu}$) in  such a way that
the background symmetry
should be restored at the perturbation level.
This is an important point in studying all black holes [7,15].
 One linearizes (8)-(11) in order to obtain the equations governing the
perturbation as

\begin{equation}
  \delta R_{\mu\nu} (h)
+ \bar \nabla_\mu \bar \nabla_\nu \phi
- \delta \Gamma^\rho_{\mu\nu} (h) \bar \nabla_\rho \bar \Phi
+ 2 \bar F_{\mu \rho} {\cal F}_\nu^{~ \rho}
- \bar F_{\mu \rho} \bar F_{\nu\alpha} h^{\rho \alpha}
 = 0,
\end{equation}

\begin{eqnarray}
&&- h^{\mu\nu} \bar \nabla_\mu \bar \nabla_\nu \bar \Phi
- \bar G^{\mu\nu} \delta \Gamma^\rho_{\mu\nu} (h) \partial_\rho \bar \Phi
+ \bar \nabla^2 \phi
- h^{\mu\nu} \partial_\mu \bar \Phi \partial_\nu \bar \Phi
+ 2 \bar G^{\mu\nu} \partial_\mu \bar \Phi \partial_\nu \phi
\nonumber   \\
&&~~~~~~~~~~~~~~~- \bar F_{\mu \nu} {\cal F}^{\mu \nu}
     + \bar F_{\mu \nu}   \bar F_{\rho}^{~\nu} h^{\mu\rho} = 0,
\end{eqnarray}

\begin{equation}
  ( \bar \nabla_\mu + \partial_\mu \bar \Phi )
    ( {\cal F}^{\mu \nu} - \bar F_\alpha^{~~\nu} h^{\alpha \mu}
                             - \bar F^{\mu}_{~~\beta} h^{\beta \nu} )
      + \bar F^{\mu \nu} (\delta \Gamma^\sigma_{\sigma \mu} (h)
      + (\partial_\mu \phi))
 = 0,
\end{equation}

\begin{equation}
\bar \nabla^2 \tilde t + \bar \nabla_\mu \bar \Phi \bar \nabla^\mu \tilde t
+ 2 \tilde t = 0,
\end{equation}
where

\begin{eqnarray}
&&\delta R_{\mu\nu} (h) = {1 \over 2} \bar \nabla_\mu \bar \nabla_\nu
h^\rho_{~\rho}
 + {1 \over 2} \bar \nabla^\rho \bar \nabla_\rho h_{\mu\nu}
 - {1 \over 2} \bar \nabla^\rho \bar \nabla_\nu h_{\rho\mu}
 - {1 \over 2} \bar \nabla^\rho \bar \nabla_\mu h_{\nu\rho},   \\
&&\delta \Gamma^\rho_{\mu\nu} (h) = {1 \over 2} \bar G^{\rho\sigma}
( \bar \nabla_\nu h_{\mu\sigma} + \bar \nabla_\mu h_{\nu\sigma} - \bar
\nabla_\sigma h_{\mu\nu} ).
\end{eqnarray}

 From (23) one can express ${\cal F}$ in terms of $\phi$ and $h$ as

\begin{equation}
{\cal F} = - Q ( \phi + h ).
\end{equation}
This means that ${\cal F}$ is no longer an independent mode.
Also from the diagonal element of (21), we have

\begin{eqnarray}
&&  \bar \nabla^2 h - 2 \bar \nabla^2_t \phi - 2 \sqrt 2 \bar G^{rr}
\partial_r h
       + \bar F^2 (h + 2 \phi) = 0,    \\
&&  \bar \nabla^2 h - 2 \bar \nabla^2_r \phi + 2 \sqrt 2 \bar G^{rr}
\partial_r h
       + \bar F^2 (h + 2 \phi) = 0.
\end{eqnarray}
Adding the above two equations leads to

\begin{equation}
\bar \nabla^2 ( h - \phi) +
  \bar F^2 (h + 2 \phi) = 0.
\end{equation}
And the off-diagonal element of (21) takes the form

\begin{equation}
 \partial_t \big\{ ( \partial_r  - \Gamma^t_{tr}) \phi + \sqrt 2  h  \big\}
   = 0.
\end{equation}
Also the dilaton equation (22) together with (27) leads to

\begin{equation}
\bar \nabla^2 \phi + 4 \sqrt 2 f \partial_r \phi + 2 \sqrt 2 (\partial_r f
+ 2 \sqrt2 f) h
+ \bar F^2 \phi = 0.
\end{equation}
{}From (31), the relation between $\phi$ and $h$ is given by
\begin{equation}
\sqrt2 h = - \partial_r \phi + {1 \over 2}{\partial_r f \over f} \phi + U(r).
\end{equation}
Here $U(r)$ is the residual gauge degrees of freedom and thus we set $U(r)
= 0$ for simplicity.
Substituting (33) into (32), we have
\begin{equation}
\bar \nabla^2 \phi + 2 \sqrt 2  \partial_r f ( h + \phi)
+ \bar F^2 \phi = 0.
\end{equation}
Calculating  (30) + 2 $\times$ (34), one finds the other equation
\begin{equation}
\bar \nabla^2 (h + \phi) + 4 \sqrt 2  \partial_r f ( h + \phi)
+ \bar F^2 ( h + 4 \phi) = 0.
\end{equation}
In the beginning we started with two fields ($h, \phi$).
However, from (30) and (35) we have four modes
($h - \phi, h + \phi, h + 2\phi, h + 4\phi$).
When $Q \ne 0$, it is not easy to find out the solutions which satisfy both
(30) and (35).
Instead, we first check whether the graviton ($h$) and  dilaton ($\phi$)
are physically
 propagating modes
in the 2D charged black hole blackground.
We consider the conventional counting of degrees of freedom.
The number of degrees of freedom for the gravitational field ($h_{\mu\nu}$) in
$d$-dimensions is $(1/2) d (d -3)$.  For a $d=4$ Schwarzschild black hole,
we obtain two degrees of freedom. These correspond to the Regge-Wheeler
mode for odd-parity perturbation
and Zerilli mode for even-parity perturbation.  We have $-1$ for $d=2$.
This means that in
two dimensions
the contribution of graviton is equal and opposite to that of a spinless
particle (dilaton).
In the 2D dilaton black hole ($ Q=0$), two graviton-dilaton modes ($h -
\phi, h + \phi$)
 are thus trivial gauge
artifacts [16,17]. In addition, a Maxwell field is introduced in the
charged 2D black hole.
 The  Maxwell field has $d-2$ physical degrees of freedom.
For $d=2$, Maxwell field has no physical degrees of freedom.  We  confirm
this from (27).
We thus insist that  graviton-dilaton, and Maxwell modes
become gauge artefacts in the charged 2D black hole.
Since these are not physically propagating modes, it is not necessary to
consider
(30) and (35).

One remaining equation that describes a physically propagating mode  is
just the tachyon equation
(24), which can be rewritten as
\begin{equation}
f^2 t''  + f f't' - [\sqrt 2 ff' - 2 f(1 - f)] t
 - {\partial^2 \over \partial t^2} t = 0,
\end{equation}
where the prime ($\prime$) denotes the derivative with respect to $r$.
In order to study the tachyonic propagation, we  transform (36)
into the form of  Schr\"odinger equation.
Introducing the  coordinate transformation
$$r\to r^* \equiv g(r),$$
then (36) can be rewritten as
\begin{equation}
f^2 g'^2 {\partial^2 \over \partial r^{*2}} t  + f \{ f g'' +  f' g'\}
{\partial \over \partial r^* }t - [\sqrt 2 ff' - 2 f (1 - f)]t
 - {\partial^2 \over \partial t^2} t = 0.
\end{equation}
Requiring that the coefficient of the linear derivative vanish, one finds
the relation
\begin{equation}
g' =  {1 \over f}.
\end{equation}
{}From this relation one can determine the explicit form of  $r^*,$
\begin{equation}
r^*= g(r) = { 1 \over \kappa_+}\log |e^{2\sqrt 2 r} - e^{2\sqrt 2 r_{+}}| -
{1 \over \kappa_- }
\log |e^{2\sqrt 2 r} -e^{2\sqrt 2 r_{-}}|,
\end{equation}
with the surface gravity at $r_{\pm}$ defined as
$$\kappa_{\pm} ={2 \sqrt2 (y_+ -y_-) \over y_{\pm}}.$$
Assuming $t( r^*,t ) \sim t_{\omega} ( r^* ) e^{-i\omega t}$,
one can cast (37) into the one-dimensional Schr\"odinger equation

\begin{equation}
\{ {\partial^2 \over \partial r^{*2}} + \omega^2 - V_T(r)\} t_{\omega} = 0,
\end{equation}
where the effective potential $V_T(r)$ is given by

\begin{equation}
V_T(r) = f(\sqrt 2 f' - 2  (1 - f)).
\end{equation}
As is shown in Fig.1, $V_T(r)$ is a double-humped barrier well ($V^{IN}_T$)
between
the Cauchy horizon and event horizon, while
it is just a potential barrier ($V^{OUT}_T$) outside the event horizon.

First we consider the region inside the black bole. It is very important to
note that inside the black
 hole the radial coordinate ($r$ or $r^*$) is timelike, whereas the time
($t$) is spacelike.
 Hence to quest the internal structure of black hole is an evolutionary
problem.
Near the horizons, the potential decreases exponentially as
\begin{equation}
 V_T^{IN}(r^*) \propto \exp (\kappa_+ r^*),~~~ r^* \to -\infty  (r \to r_+)
\end{equation}
and
\begin{equation}
 V_T^{IN}(r^*) \propto \exp (-\kappa_- r^*),~~~ r^* \to \infty  (r \to r_-).
\end{equation}
It is useful to introduce the null coordinates ($ v = r^* + t, u = r^* -
t$) to describe the
inner structure of the charged black hole. In these coordinates the metric
is given by $ds^2 = fdvdu$.
As is shown in Fig.2, the Cauchy horizon $r=r_-$ consists of two branches
(the right with $v=\infty$ and the left with $u=\infty$).
In order to find the energy density of the tachyon measured by a freely
falling observer (FFO)
with two-velocity $U^{\mu} (U^{\mu}U_{\mu}=-1)$,
we have to consider the boundary conditions. Initially the tachyonic mode
falls into the hole from
the exterior region.   After solving the equation near the horizons
\begin{equation}
\{ {\partial^2 \over \partial r^{*2}} + \omega^2 \} t^{IN}_{\omega} = 0,
\end{equation}
we have the ingoing wave near the event horizon $(r_{+})$
\begin{equation}
t^{IN}_{\omega} e^{-i\omega t}\mid_{r_+} = T^{IN}(\omega) \exp\{-i\omega(t
+ r^*)\}.
\end{equation}
On the other hand, the boundary condition near the Cauchy horizon is
\begin{equation}
t^{IN}_{\omega} e^{-i\omega t}\mid_{r_-} = \exp\{-i\omega(t + r^*)\} +
R^{IN}(\omega)
\exp\{-i\omega(t - r^*)\},
\end{equation}
where the first term refers the ingoing mode into the left branch with
$u=\infty$, while the second
denotes the backscattered mode into the right branch with $v=\infty$.
Here $T^{IN}(\omega)$ and $R^{IN}(\omega)$ are the transmission and
reflection amplitudes for given
mode $\omega$. We need the total tachyonic function $(\tilde t^{IN}(r^*,t)$
in (20)) to
obtain the energy density.
This is given by  the Fourier integral transfrom over the frequency $\omega$
\begin{equation}
\tilde t^{IN}(r^*,t)= \int e^{-\sqrt2 r}t^{IN}_{\omega} e^{-i\omega t}
a(\omega) d\omega
\end{equation}
with the mode constant $a(\omega)$.
Considering the boundary condition near the Cauchy horizon, this takes the form
\begin{equation}
\tilde t^{IN}(u,v)\mid_{r_-} \sim e^{-\sqrt2 r}[t^{-}(u) + t^{+}(v)].
\end{equation}
{}From (12) and (20), the energy density measured by a FFO is dominated by [6]
\begin{equation}
\rho = T^t_{\alpha\beta}U^\alpha U^\beta \sim |U^{\alpha} \tilde
t^{IN},_{\alpha}|^2 \sim |U^{\alpha}t^{\pm},_{\alpha}|^2.
\end{equation}
When  a FFO1 crosses the left branch of the Cauchy horizon, one has
\begin{equation}
   U^{\alpha} t^{-},_{\alpha} \propto t^{- \prime}(u) \exp( {\kappa_+ u
\over 2}),
\end{equation}
where the prime means the differentiation with respect to the given argument.
In order to calculate $t^{- \prime}(u)$, we consider the deviation from the
wave
($t^{IN}_{\omega}= \exp(-i\omega r^*)$)
treating $V_T^{IN}(r^*)$ in (43) as the infinitesimal perturbation.
Following Ref.[6], we find $t^{- \prime}(u) \propto \exp (-{\kappa_+ u
\over 2})$ as $u \to \infty$.
Therefore this wave gives a finite energy density at the left Cauchy horizon.
On the other hand, the energy density measured by a FFO2 who crosses the
right($v \to \infty$)
horizon is proportional to the square of
\begin{equation}
U^{\alpha} t^{+},_{\alpha} \propto t^{+ \prime}(v) \exp( {\kappa_- v \over 2}).
\end{equation}
Substituting the form of $t^{+ \prime}(v) \propto \exp (-{\kappa_+ v \over
2})$ into (51)
together with $\kappa_- - \kappa_+ > 0$ leads to a divergent
energy density on the right Cauchy horizon. The monochromatic tachyon waves
with small amplitude
and purely ingoing near the event horizon develop the infinite energy
density near Cauchy horizon.
This corresponds to the blueshift of tachyon. Further
this means that the Cauchy horizon of the 2D charged black hole is unstable
to the physical perturbations.

Now let us consider the same problem outside the black hole.
Note that the last term (${Q^2 \over 8}e^{-4\sqrt 2 r}$) of $f$ in (14)
decreases faster than the second ($e^{-2 \sqrt 2 r}$), as $r$ increases.
Outside the black hole, we then immediately  recover the 2D dilaton black
hole background as
\begin{equation}
\bar \Phi = 2 \sqrt 2 r,~~~ \bar F_{tr} \approx 0,~~~ \bar T = 0,
{}~~~ \bar G_{\mu\nu} \approx
 \left(  \begin{array}{cc} - \tilde f & 0  \\
                             0 & \tilde f^{-1}   \end{array}   \right)
\end{equation}
with
\begin{equation}
\tilde f = 1 -  e^{- 2 \sqrt 2 r}.
\end{equation}
This corresponds to  $Q = 0$ case.
The difference is that the position of event horizon is shifted from $r_+ =
- 0.056$ (for $Q\not=0$)
to $ r_+ = 0$.
Actually $\tilde V_T^{OUT}$ in Fig. 3( potential for the 2D dilaton black
hole) is approximately
a copy of the
right barrier ($V_T^{OUT})$ in Fig.1 except the shift of $r_+$ and scaling.
These differences are not important to our considerations
outside the black hole. For a simplicity, we use  $\tilde V_T^{OUT}$
(instead of $V_T^{OUT}$)
to investigate the exterior region.  Here we briefly study
how string modes propagate outside the 2D dilaton black hole [15].
Introducing $\tilde r^* = \tilde g(r)$, we  have $\tilde g'\tilde f=1$.
Explicitly, one can find
the form of $\tilde g(r)$ as

\begin{equation}
 \tilde g(r) = r + {1 \over 2 \sqrt 2} \ln (1 - e^{- 2 \sqrt 2 r}).
\end{equation}
Note that $\tilde r^*$ ranges from $- \infty$ to $+ \infty$, while $r$ ranges
from the event horizon of the black hole ($ r_+ = 0$) to $+ \infty$.
We can visualize the 2D dilaton black hole as presenting a potential
barrier (well) to the oncoming waves (for example, $t, h + \phi, h - \phi$).
First let us discuss the tachyonic propagations outside the black hole.
Assuming $t^{OUT}( r^*,t ) \sim t^{OUT}_\omega ( r^* ) e^{-i \omega t}$,
we find the equation for the tachyonic mode from (36)
\begin{equation}
[ {\partial^2 \over \partial r^{*2}} + \omega^2 - \tilde V^{OUT}_T ]
L^{OUT} = 0,
\end{equation}
where $\tilde V^{OUT}_T$ is given by

\begin{equation}
  \tilde V^{OUT}_T =  \tilde f(\sqrt 2 \tilde f' - 2  (1 - \tilde f)) =
2 e^{-2 \sqrt{2} r} ( 1 - e^{-2 \sqrt{2} r }).
\end{equation}
The scattering of tachyon by (56) was discussed in [15] and is shown in Fig.2.
For graviton-dilaton modes, (30) and (35) reduce to
\begin{equation}
\bar \nabla^2 (h - \phi) = 0,
\end{equation}

\begin{equation}
\bar \nabla^2 (h + \phi) + 4 \sqrt 2  \partial_r \tilde f ( h + \phi) = 0.
\end{equation}
Here we obtain two modes
$(h - \phi)$ and $(h + \phi)$.
Defining $H \equiv h - \phi$ and
considering the trial solution of the form

\begin{equation}
H^{OUT}(r^*,t) \sim H^{OUT}_k(r^*) e^{-i k t},
\end{equation}
 we have the free field equation from (57)

\begin{equation}
[ {\partial^2 \over \partial r^{*2} } + k^2 ] H^{OUT}_k( r^* ) = 0.
\end{equation}
The other graviton-dilaton mode ($J \equiv h + \phi$) is also
 given by
the one-dimensional differential equation. Considering $J^{OUT}( r^*,t) \sim
 J^{OUT}_k (r^*) e^{- i k t}$ , from (58) we obtain

\begin{equation}
[ {d^2 \over d r^{*2}} + ( k^2 - V^{OUT}_J ) ] J^{OUT}_k = 0,
\end{equation}
where the potential well is given by

\begin{equation}
V^{OUT}_J =  - 16e^{-2 \sqrt{2} r} ( 1 - e^{-2 \sqrt{2} r})
    = { - 4 \over (\cosh \sqrt{2} r^*)^2 }.
\end{equation}

Outside 2D dilaton black hole, we find the well-known Schr\"odinger problem
with the constant
energy ($E=k^2$) : a potential
barrier (56) for the tachyonic mode, a potential well (62) for one
graviton-dilaton mode
$(h + \phi)$ and no potential for the other graviton-dilaton mode
$(h - \phi)$. It is emphasized that the potential well for $J = h + \phi$ is
obviously a new feature of $d=2$ black hole.   From (61), one finds an
 exponentially growing mode with time ($ e^{-i k t}, k=i\alpha$).  Naively
this means that the event
horizon is unstable. However, according to [16],
 $(h + \phi)$ and  $(h - \phi)$ are not the physical degrees of freedom and
nothing but gauge
artifacts. Further it is explicitly shown that this exponentially growing
mode with time
can be removed by the coordinate transformation (just a translation) [17].
 The tachyonic mode is also the  physically propagating one outside  the
black hole.
The stability should be based on the physical degree of freedom.
With the potential barrier (56), one cannot find
the bound state solutions which lead to the exponentially growing modes.
Therefore the event (outer) horizon of 2D charged black hole is stable.

In summary, all string fields except the tachyon are non-propagating in the 2D
charged black hole with double horizons.  Only the tachyonic mode is
a physically propagating one  both inside and outside  the black hole.
The crucial problem has to do with the inner
(Cauchy) horizon of the 2D charged black hole.   Inside the black hole the
radial coordinate
($r$ or $r^*$) is timelike, whereas the time $(t)$ is spacelike.
Hence to investigate the internal structure of
 black hole is an evolutionary problem. This horizon is believed to be unstable
under tachyonic  perturbation because it is a surface where the infalling
matter got
infinitely blueshifted [2-8]. On the other hand, outside the black hole we
have the conventional
scattering problem with the potntial barrier. The outer (event)
horizon of 2D charged black hole is stable.
\acknowledgments

This work was supported in part by the Basic Science Research Institute
Program,
Ministry of Education, Project NO. BSRI-95-2441
and by NONDIRECTED RESEARCH FUND, Korea Research Foundation, 1994.

\figure{ Fig.1 : The graph of the effective potential of tachyon  ($
V_T(r)$). This takes the
   double-humped barrier well ($V_T^{IN}$) inside the black hole, while it
takes a simple potential
 barrier ($V_T^{OUT}$)
   outside the black hole. The event
  horizon is at $r_{+}= -0.056$ and the Cauchy  horizon is at $r_{-}= -0.679$.

\figure{ Fig.2 : Conformal diagram of a portion of the 2D charged black
hole space-time.
Two observers
  are  shown falling through $r=r_+$ into the interior region and then
through the Cauchy horizon
at $r=r_-$. FFO1 (FFO2) cross  the left (right) branches of $r=r_-$. An
incident wave is scattered
from the potential ($V_T^{OUT}$), then proceeds into the interior region
where further scattering
by  $V_T^{IN}$ occurs. The scattered wave will be rescattered into hole, to
give a tail with a
power-law (in time) decay. The energy density near the right branch due to
this tail decays
sufficiently slowly so that infinite energy densities are developed. These
are measured by FFO2. }

\figure{ Fig.3 : The graph of the  potential barrier of tachyon outside 2D
dilaton black hole
 ($ \tilde V_T^{OUT}$).
  This is approximately a  copy of right barrier ($V_T^{OUT}$) in Fig.1.
The apparent difference is
due to a shifting from $r_{+}= -0.056 (Q \not= 0)$ to $r_{+}= 0$ and
scaling. For a simplicity we
 use this potential to investigate the exterior region. The
  asymptotically flat region is located  at $r=\infty$.}
\end{document}